\documentclass[%
 reprint,
 amsmath,amssymb,
 aps
]{revtex4-2}

\usepackage{dcolumn}
\usepackage{bm}
\usepackage{xcolor}
\usepackage{textcomp}
\usepackage{comment}
\usepackage{diagbox} 
\usepackage{graphicx} 
\graphicspath{{Figures/}} 
\begin{document}


\title{Topological charge of finite-size photonic crystal modes}




\author{Zhixin Wang}
\email{Corresponding author: zhixwang@phys.ethz.ch}

\author{Yong Liang}
\author{Mattias Beck}
\author{Giacomo Scalari}
\author{J\'er\^ome Faist}

\affiliation{ETH Z\"urich, Institute of Quantum Electronics, Auguste-Piccard-Hof 1, Z\"urich 8093, Switzerland}



\date{\today}

\begin{abstract}
Topological charges are the winding numbers of polarization vectors around the vortex centers of far-field radiation. In this work, the topological charge of photonic crystal modes is theoretically analyzed using an envelope function approach. A group of modes is discovered with unique polarization properties dictated by their non-trivial envelope functions. Experimentally, lasing operation on such mode is demonstrated in an electrically pumped mid-infrared photonic crystal surface-emitting laser with high slope efficiency. The topological charge is directly observed from the polarization properties of single-mode laser emission.
\end{abstract}


\maketitle

The past decade witnessed a boom of research in the field of topological photonics. This topic originates from the recent discoveries in solid-state materials, especially the topological insulator \cite{hasan2010colloquium,qi2011topological} and the quantum Hall effect \cite{laughlin1981quantized,haldane1988model,hansson2017quantum}. Topology brings a new perspective to the classification of photonic systems \cite{raghu2008analogs,wang2009observation,lu2014topological,ozawa2019topological}. 

The key feature of topological protection can be implemented into a photonic system with several different approaches. By introducing a pseudo magnetic gauge field, a system of coupled ring resonators exhibit topological robustness \cite{hafezi2011robust,hafezi2013imaging} and enable topological insulator lasers \cite{harari2018topological,bandres2018topological}. Dynamic modulation of ring resonators generates topological edges in the synthetic space \cite{yuan2016photonic,yuan2018synthetic}. By direct analogy to the periodic electron system, the photonic crystal \cite{joannopoulos_photonic_2011} naturally serves as an ideal platform for observing topological invariants \cite{lee2012observation,hsu_observation_2013,zhen_spawning_2015,zhou2018observation}, such as the Chern number \cite{ma2016all,gao2018topologically} and the topological charge \cite{zhen_topological_2014,zhou2018observation,jin2019topologically}. The topological charge is defined as the winding number of the radiating polarization vector around a vortex center, where no field vector can be assigned \cite{zhen_topological_2014}. Such a singular point can be created either at or off $\Gamma$ point \cite{hsu_observation_2013}. 
The concept of topological charge is not limited to the area of photonics. Similar phenomena can be explored also in solid-state physics, such as the magnetic skyrmions \cite{heinze2011spontaneous}. In fact, the topological charge is closely associated with the polarization vectors of far-field radiation from photonic crystal surface-emitting lasers \cite{zhen_topological_2014}, which have been well-developed for high-power and high-brightness laser applications \cite{hirose2014watt,yoshida2019double}. 

Photonic crystal surface-emitting lasers \cite{miyai2006photonics} are typically designed at the second-order $\Gamma$ ($\Gamma_2$) point of a photonic band structure, where lasing action naturally occurs at symmetry-protected band-edge modes \cite{wang2019large,liang2019room}. Real devices have finite dimensions, which can be simulated in theory by the exact diagonalization of all mesh elements in three dimensions (3D). However, brute-force techniques are unattractive, as the 3D calculations rapidly become intractable for large devices, and they do not typically yield physical insights. Instead, the envelope functions are introduced to the wave-functions of finite periodic systems as an approximation approach \cite{bastard1990wave}. 

Previously, the photonic band structure alone was considered to determine the topological charge of a photonic crystal device. The role of near-field envelopes is empirically neglected when predicting the far-field patterns. However, in general this is not correct. A full symmetry analysis must include both the periodic waves and the envelope functions. Take electron states in a symmetric quantum well for example. Determined by the interaction Hamiltonian $H_{int} = - \frac{e}{m}\bm{A \cdot p}$, optical transitions are allowed only between states with the opposite parities. Interband transitions occur between states with the opposite periodic function parities and the same envelope parities, whereas intersubband transitions are allowed between states with the opposite envelope parities and the same periodic function parities. In this work, we explicitly bring the envelope functions into the analysis of the topological charge of photonic crystal modes.

\begin{figure*}[!htbp]
\centering\includegraphics[width=\linewidth]{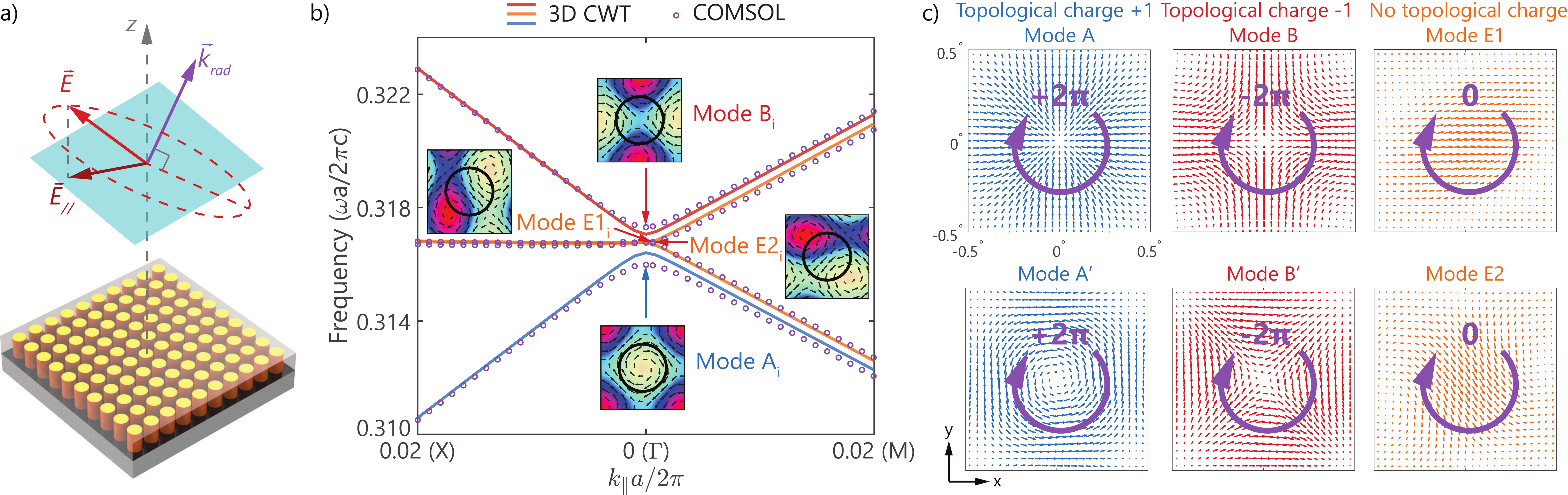}
\caption{(a) Schematic drawing of the photonic crystal structure and the radiating field. (b) Photonic band structure calculated by both coupled wave theory and COMSOL. Here the pillar filling factor is 0.25. The band-edge mode patterns in one unit cell are shown in the insets, where the colormaps depict the $E_z$ component, and the vector maps show the in-plane magnetic field $H_x$, $H_y$. Mode $A_i$ is $C_4$-even, mode $B_i$ is $C_4$-odd, and modes $E1_i$, $E2_i$ are both $C_2$-odd. (c) Far-field beam vector ($E_{\parallel}$) patterns of the fundamental finite-size photonic crystal modes calculated by the coupled wave theory. The pillar filling factor of the photonic crystal is 0.25. The device size is 400 periods in each in-plane dimension. Following a closed loop around the center point, the polarization vectors of modes $A$ and $A'$ rotate by an angle of $+2\pi$, and thus their topological charges are +1. By the same token, the topological charges of modes $B$, $B'$ are -1. The winding number for modes $E1$ and $E2$ is 0. These two modes possess no topological charge in the center point.}\label{fig:1}
\end{figure*}

Consider a two-dimensional square-lattice photonic crystal with cylindrical pillars [Fig. \ref{fig:1}(a)]. Such a structure is invariant under the $C_2$ and $C_4$ operators, which enact a $180^{\circ}$ and $90^{\circ}$ rotation around the out-of-plane ($z$) axis, respectively. Figure \ref{fig:1}(b) shows the photonic band structure of transverse magnetic (TM) modes calculated by the coupled wave theory \cite{yin2017analytical,wang2017analytical} and the finite-element method (COMSOL Multiphysics), with the pillar filling factor of 0.25. As indicated in the insets of Fig. \ref{fig:1}(b),  the four band-edge modes at the $\Gamma_2$ point possess different symmetries. The electromagnetic field of the monopole mode $A_i$ (the subscript $_i$ means infinite) is even under both $C_2$ and $C_4$ rotations. The quadrupole mode $B_i$ is $C_2$-even and $C_4$-odd. The degenerate dipole modes $E1_i$ and $E2_i$ are $C_2$-odd. The vertical radiations of modes $A_i$ and $B_i$ are topologically forbidden, since the field profile of a plane wave is $C_2$-odd. Consequently, topological charges are created at $k_{\parallel} = 0$, where the projected average fields [$E_{\parallel}$ in Fig. \ref{fig:1}(a)] are zero. 


Similar to a confined electron system, the wave-functions in a finite photonic crystal structure can be approximated as periodic Bloch functions with envelope functions. In the coupled wave theory, this treatment is mathematically interpreted as the solutions to a semi-analytical eigen-equation \cite{liang_three-dimensional_2012,wang2017analytical}:
\begin{equation}
f \vec{V} = \mathbf{C_{infinite}} \vec{V} + \mathbf{C_{finite}} \vec{V}
\label{equ:master}
\end{equation}
where $f$ is the normalized eigen-frequency, $\vec{V}$ denotes the fundamental Bloch waves [$(k_x, k_y) = (0,\pm \frac{2\pi}{a}) $and $(\pm \frac{2\pi}{a},0)$] in all locations and directions, $\mathbf{C_{infinite}}$ is a matrix representing the coupling of the Bloch waves due to the Bragg reflection, and $\mathbf{C_{finite}}$ is a matrix describing the coupling induced by the in-plane boundaries. In solutions to Eq. \ref{equ:master}, the periodic functions perform as a basis set, and the envelope functions behave as the combination coefficients that vary in different locations of the device. However, the choice of periodic basis is not limited to the fundamental Bloch waves. Instead, the four band-edge modes with distinct symmetry properties build up a new periodic basis to describe a finite photonic crystal mode $M_f$:
\begin{equation}
M_{f} = [N_{A_i}^{M_f}(x,y), N_{B_i}^{M_f}(x,y), N_{E1_i}^{M_f}(x,y), N_{E2_i}^{M_f}(x,y)] \vec{V}_{PhC}
\label{equ:band-edge-basis}
\end{equation}
where $\vec{V}_{PhC} = [A_i, B_i, E1_i, E2_i]^T$ is the vector of band-edge modes and $N(x,y)$ is the envelope function for each term. In this basis, the rotational symmetry of the finite mode can be predicted by the shapes of envelope functions. The band-edge basis is physically analogous to the $\bm{k \cdot p}$ method of the solid-state physics where the off-$\Gamma$ states are described by the wavefunctions at $k = 0$ \cite{kittel1963quantum}. See Appendix Sec. A for more details on the basis transformation \cite{SUPPLEMENT}.

Frequencies, intensity profiles and far-field patterns of the photonic crystal modes are obtained by solving Eq. \ref{equ:master}. Among all the finite TM modes, six fundamental ones are found to possess the smallest in-plane wavevector ($k_{\parallel}$) components and the lowest cavity losses. Their far-field patterns are shown in Fig. \ref{fig:1}(c). Here the calculation is conducted on a 4-layer photonic crystal device with a square in-plane boundary. The device has 400 periods in each in-plane dimension. On the $x$,$y$ axes, modes $A$ and $B$ are radially polarized, which is predicted by the symmetry analysis of infinite TM modes (see Appendix Sec. B \cite{SUPPLEMENT}). No field vector can be observed in the center points of the far-field patterns for modes $A$, $A'$, $B$ and $B'$. In a counter-clockwise loop around the pattern centers in Fig. \ref{fig:1}(c), the field vectors of modes $A$, $A'$ ($B$, $B'$) rotate by a phase of $+2\pi$ ($-2\pi$), leading to topological charges of +1 (-1). Modes $E1$ and $E2$ have no topological charge in their center points. Although mode $A'$ ($B'$) shares the same topological charge as mode $A$ ($B$), their polarization vectors are indeed orthogonal to each other in almost every direction. Surprisingly, both modes $A'$ and $B'$ are azimuthally polarized on the $x$,$y$ axes, which is contradictory to all previous reports on TM modes.

\begin{figure}[!htbp]
\centering\includegraphics[width=\linewidth]{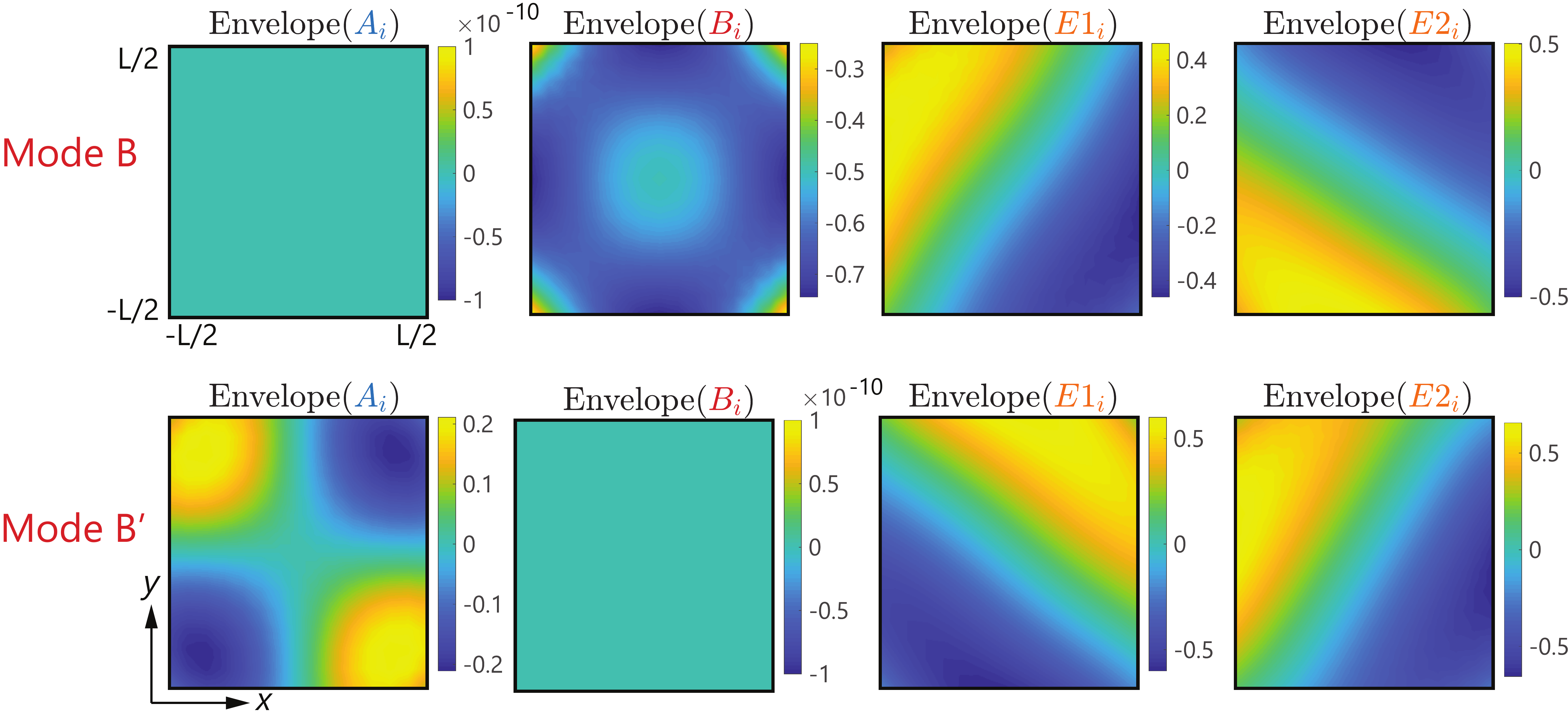}
\caption{Envelope functions of modes $B$ and $B'$ in the basis of photonic crystal band-edge modes. The values are in arbitrary units. The $A_i$ envelope of mode $B$ and the $B_i$ envelope of mode $B'$ both equal to zero, as shown in the first figure on the upper row, and the second figure in the lower row.}\label{fig:2}
\end{figure}

An intrinsic difference between modes $A'$, $B'$ and the other four modes is the near-field intensity profiles. With large device dimensions where the in-plane confinement is sufficient, the intensity profiles of modes $A$, $B$, $E1$ and $E2$ are convex in shape with anti-nodes in the device center. The near-field intensities of modes $A'$ and $B'$, however, are concave in shape and are pinned to zero at the device center regardless of the device dimensions. See Appendix Sec. C for more details \cite{SUPPLEMENT}.

\begin{table}[!htbp]
\begin{ruledtabular}
\begin{tabular}{c|cccc}
 & $A_i$          & $B_i$       & $E1_i$       & $E2_i$     \\
 & ($C_4$-even)      & ($C_4$-odd) & ($C_2$-odd)  & ($C_2$-odd)\\
\hline 
Mode $A$  & $C_4$-even & 0         & $C_2$-odd & $C_2$-odd  \\
Mode $A'$ & 0         & $C_4$-odd  & $C_2$-odd & $C_2$-odd  \\
Mode $B$  & 0         & $C_4$-even & $C_2$-odd & $C_2$-odd  \\
Mode $B'$ & $C_4$-odd & 0          & $C_2$-odd & $C_2$-odd  
\end{tabular}
\end{ruledtabular}
\caption{%
The summarized $C_4$ and $C_2$ symmetries of the envelopes functions for the photonic crystal modes $A$, $A'$, $B$ and $B'$. The envelope functions are calculated on the basis of the band-edge modes, which possess distinct symmetry properties individually. The topological charge of the photonic crystal mode can be predicted by the symmetries of the band-edge basis terms and their envelope functions.
}\label{tab:symmetry}
\end{table}

We further investigate the near-field envelope functions, by comparing two modes $B$ and $B'$, which hold the same topological charge. Their envelope functions on the band-edge mode basis (Eq. \ref{equ:band-edge-basis}) is shown in Fig. \ref{fig:2}. As depicted in Fig. \ref{fig:1}(b), the band-edge terms $A_i$, $B_i$ are $C_2$-even, and $E1_i$, $E2_i$ are $C_2$-odd. As explained above, topological charges at the radiation center require the mode to be $C_2$-even, which means the envelope functions of terms $A_i$, $B_i$ must be $C_2$-even or zero, and the envelopes of $E1_i$, $E2_i$ must be $C_2$-odd or zero. This is exactly the case for both modes $B$ and $B'$ in Fig. \ref{fig:2}. Moreover, mode $B$ has a $C_4$-even envelope for the $C_4$-odd component $B_i$, and mode $B'$ has a $C_4$-odd envelope for the $C_4$-even component $A_i$. In both cases, the periodic term and the envelope function have the opposite $C_4$ symmetries, resulting in $C_4$-odd finite modes. Hence, the value of their topological charges is -1. Mode $B$ is composed of band-edge modes $B_i$, $E1_i$, $E2_i$, and mode $B'$ by $A_i$, $E1_i$, $E2_i$. Although mode $B'$ shares the same topological charge as the mode $B$, the contribution from $B_i$ to mode $B'$ is zero. It is the non-trivial envelope function that causes such unexpected results. The +1 topological charges of modes $A$ and $A'$ can be analyzed in the same manner, of which the field is $C_4$-even. The full envelope functions of all the six fundamental finite modes [Fig. \ref{fig:1}(c)] on the band-edge basis are shown in Appendix Sec. D \cite{SUPPLEMENT}. The envelope symmetries of modes $A$, $A'$, $B$ and $B'$ are summarized in Tab. \ref{tab:symmetry}. Besides, the envelope function approach can also predict the unique polarization of modes $A'$ and $B'$. 
See Appendix Sec. E for more details on the envelope analysis \cite{SUPPLEMENT}.

\begin{figure*}[!htbp]
\centering\includegraphics[width=\linewidth]{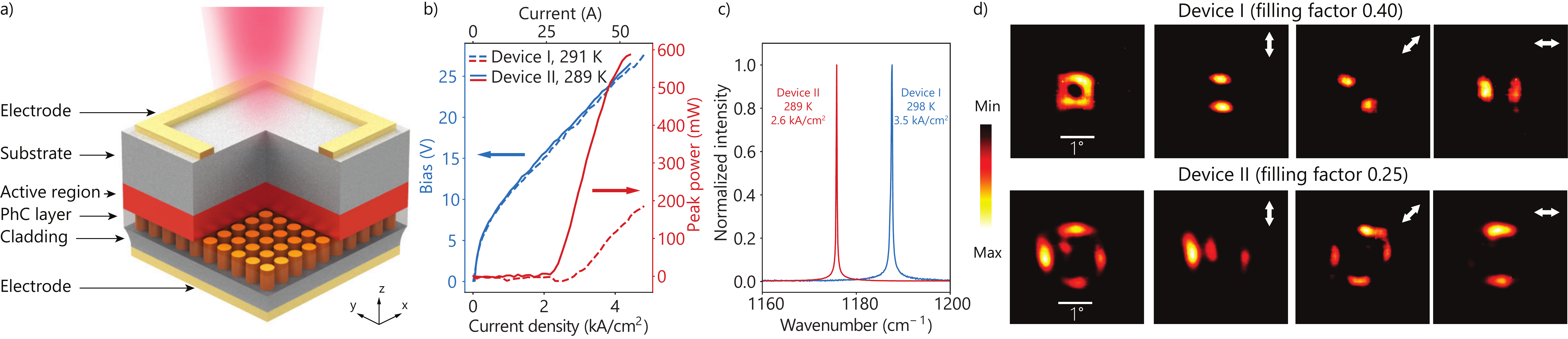}
\caption{(a) Schematic design of the surface-emitting photonic crystal lasers. The photonic crystal layer consists of circular InGaAs pillars arranged in a square lattice, surrounded by Si-doped InP. The laser is around 1.1 mm in each in-plane dimension. The emission window is opened at the substrate side. (b) LIV characterization and (c) lasing spectra of two types of lasers, Device I (pillar filling factor 0.40 / lattice-constant 2.68 \textmu m) and Device II (pillar filling factor 0.25 / lattice-constant 2.71 \textmu m). Both measurements are made with 52 ns pulses at the repetition frequency of 9.615 kHz, with the power collected through surface emission. (d) Far-field patterns and the polarization characteristics taken under the same conditions as in (c). Device I is measured at 3.2 kA/cm$^2$, and device II is measured at 2.5 kA/cm$^2$. The white arrows represent the direction of the polarizer in each measurement. Both devices exhibit a narrow beam divergence of less than $2^{\circ}$, expected for large-area single-mode oscillation. }\label{fig:3}
\end{figure*}

To prove the existence of such non-trivial envelope functions, we fabricated photonic crystal quantum cascade lasers (QCLs) \cite{colombelli2003quantum,mahler2011photonic,liang2019room} with a large emission area of  $1.1 \times 1.1$ mm$^2$. The emission wavelength is around 8.5 \textmu m and the fabrication details are described in our previous publication \cite{liang2019room}. The square-lattice photonic crystal layer consists of circular-shaped InGaAs pillars surrounded by Si-doped InP, as shown schematically in Fig. \ref{fig:3}(a). 
We designed two types of photonic crystal QCLs: device I and device II, which have different pillar sizes and lattice constants.
The candidate lasing mode of device I is designed to be mode $B$, which features the lowest cavity loss and a high optical overlap factor with the active region. By contrast, device II is targeted at mode $B'$.
See Appendix Sec. F for more details on the cavity losses of the modes \cite{SUPPLEMENT}.

The LIV characteristics and the lasing spectra of the two lasers are shown in Figs. \ref{fig:3}(b) and (c), where both lasers perform single-mode operation at room temperature well above the lasing threshold. The slope efficiency of device II [solid lines in Fig. \ref{fig:3}(b)] is almost three times as that of device I [dashed lines in Fig. \ref{fig:3}(b)]. 
Since both devices are fabricated from the same wafer, such a distinct difference implies the possibility that their lasing modes are not the same.

To identify the lasing modes, the surface-emitting far-field patterns and the polarization characteristics are measured, as shown in Fig. \ref{fig:3}(d). In both cases, the beam patterns are dark in the center, indicating the singular vortex centers. From the winding number of the electric-fields around the center, the observed topological charges are both -1. The polarization of device I is radial on the $x,y$ axes, and is azimuthal on the diagonal axis, while the polarization of device II by contrast is orthogonal to device I in all cases. In agreement with the design target, Fig. \ref{fig:1}(c) allows us to identify modes $B$ and $B'$ to be the lasing modes of device I and device II, respectively.

The switch of the lasing mode is evidenced by the significant change in the slope efficiency.
As shown in Appendix Sec. C, the field of mode $B$ concentrates in the device center, whereas the field of mode $B'$ concentrates on the device boundaries. Since the electrical pumping is non-uniform, mode $B'$ experiences a higher effective gain than mode $B$.
Energy concentration near the edges also makes the lasing mode sensitive to boundary conditions. As seen in Fig. \ref{fig:3}(d), the beam pattern of mode $B'$ is less symmetrical compared to the pattern of mode $B$. Nevertheless, the topological charge of -1 is clearly observed.


In this work, the topological charge of photonic crystal modes is theoretically analyzed with an envelope function approach. A class of modes is found to exhibit unique polarization characteristics determined by their non-trivial envelope functions. Experimentally, we demonstrate an electrically injected mid-infrared photonic crystal laser that operates on such a non-trivial mode, with three times the slope efficiency compared to that obtained with standard designs. We derive the topological charges directly from the measured polarization profiles of the surface-emitting beams.
This work bridges the research of topological photonics and photonic crystal lasers. Inspired by the electronic quantum well system, the envelope function analysis is introduced to the study of topological charges, leading to unique modes that significantly enhance the slope efficiency of the photonic crystal lasers. We emphasize that, although the experiment is performed with mid-infrared QCLs, the physics of topological charges and the essential role of envelope functions are generally valid for all photonic crystal systems.

\section*{Funding Information}
H2020 European Research Council Consolidator Grant CHIC (724344); FP7 People: Marie-Curie Actions (FEL-27 14-2).





\pagebreak
\begin{center}
\textbf{\large Appendix}
\end{center}
\setcounter{equation}{0}
\setcounter{figure}{0}
\setcounter{table}{0}
\setcounter{page}{1}
\makeatletter
\renewcommand{\theequation}{A\arabic{equation}}
\renewcommand{\thefigure}{A\arabic{figure}}




\subsection{Expansion of finite-size photonic crystal mode on the band-edge basis}
\label{supsec:band-edge-basis}
In an infinitely periodic photonic crystal strucutre, solutions to Eq. 1 in the main text are the four band-edge modes, $A_i$, $B_i$, $E1_i$ and $E2_i$ \cite{wang2017analytical}. Take mode $A_i$ as an example, the field of $A_i$ can be written as a linear combination of four Bloch waves: 
\begin{equation}
A_i = [R_x^{A_i}, S_x^{A_i}, R_y^{A_i}, S_y^{A_i}] \begin{pmatrix} e^{-i\beta_0 x} \\ e^{+i\beta_0 x} \\ e^{-i\beta_0 y} \\ e^{+i\beta_0 y} \end{pmatrix}
\end{equation}
where $R_x^{A_i}$, $S_x^{A_i}$, $R_y^{A_i}$ and $S_y^{A_i}$ are the corresponding coefficients of the Bloch wave terms.

The four band-edge modes construct a new vector $\vec{V}_{PhC}$:
\begin{equation}
\vec{V}_{PhC} = 
\begin{pmatrix} A_i \\ B_i \\ E1_i \\ E2_i \end{pmatrix} = \mathbf{T_0}
\begin{pmatrix} e^{-i\beta_0 x} \\ e^{+i\beta_0 x} \\ e^{-i\beta_0 y} \\ e^{+i\beta_0 y} \end{pmatrix} 
 = \mathbf{T_0} \vec{V}_{Bloch}
\label{equ:band-edge}
\end{equation}
where each row of the transfer matrix $\mathbf{T_0}$ consists of the Bloch wave coefficients for each band-edge mode. For example, the first row of $\mathbf{T_0}$ equals to $[R_x^{A_i}, S_x^{A_i}, R_y^{A_i}, S_y^{A_i}] $.

Similarly, in the finite-size case, the field of a photonic crystal mode $M_f$ is described by the envelopes of the Bloch waves:
\begin{equation}
M_{f} = [R_x^{M_f}(x,y), S_x^{M_f}(x,y), R_y^{M_f}(x,y), S_y^{M_f}(x,y)] \vec{V}_{Bloch}
\label{equ:finite-mode}
\end{equation}
where $R_x^{M_f}(x,y)$, $S_x^{M_f}(x,y)$, $R_y^{M_f}(x,y)$ and $S_y^{M_f}(x,y)$ are the corresponding envelope functions of the Bloch wave terms.

The periodic basis can be transformed from the typical Bloch waves to the band-edge modes by substituting $\vec{V}_{Bloch}$ in Eq. \ref{equ:finite-mode} with the reverse of Eq. \ref{equ:band-edge}:
\begin{equation}
\vec{V}_{Bloch} = \mathbf{T^{-1}_0} \vec{V}_{PhC}
\end{equation}

Therefore, the field of finite mode $M_f$ can be expanded in the band-edge basis as:
\begin{equation}
M_{f} = [N_{A_i}^{M_f}(x,y), N_{B_i}^{M_f}(x,y), N_{E1_i}^{M_f}(x,y), N_{E2_i}^{M_f}(x,y)] \vec{V}_{PhC}
\end{equation}
where
\begin{equation*}
\begin{split}
& [N_{A_i}^{M_f}(x,y), N_{B_i}^{M_f}(x,y), N_{E1_i}^{M_f}(x,y), N_{E2_i}^{M_f}(x,y)] \\
= & [R_x^{M_f}(x,y), S_x^{M_f}(x,y), R_y^{M_f}(x,y), S_y^{M_f}(x,y)] \mathbf{T^{-1}_0}
\end{split}
\end{equation*}

\subsection{Polarization of photonic crystal modes predicted from an infinite system: conventional wisdom}
\label{supsec:inf_pol}
\begin{figure}[!htbp]
\centering\includegraphics[width=0.75\linewidth]{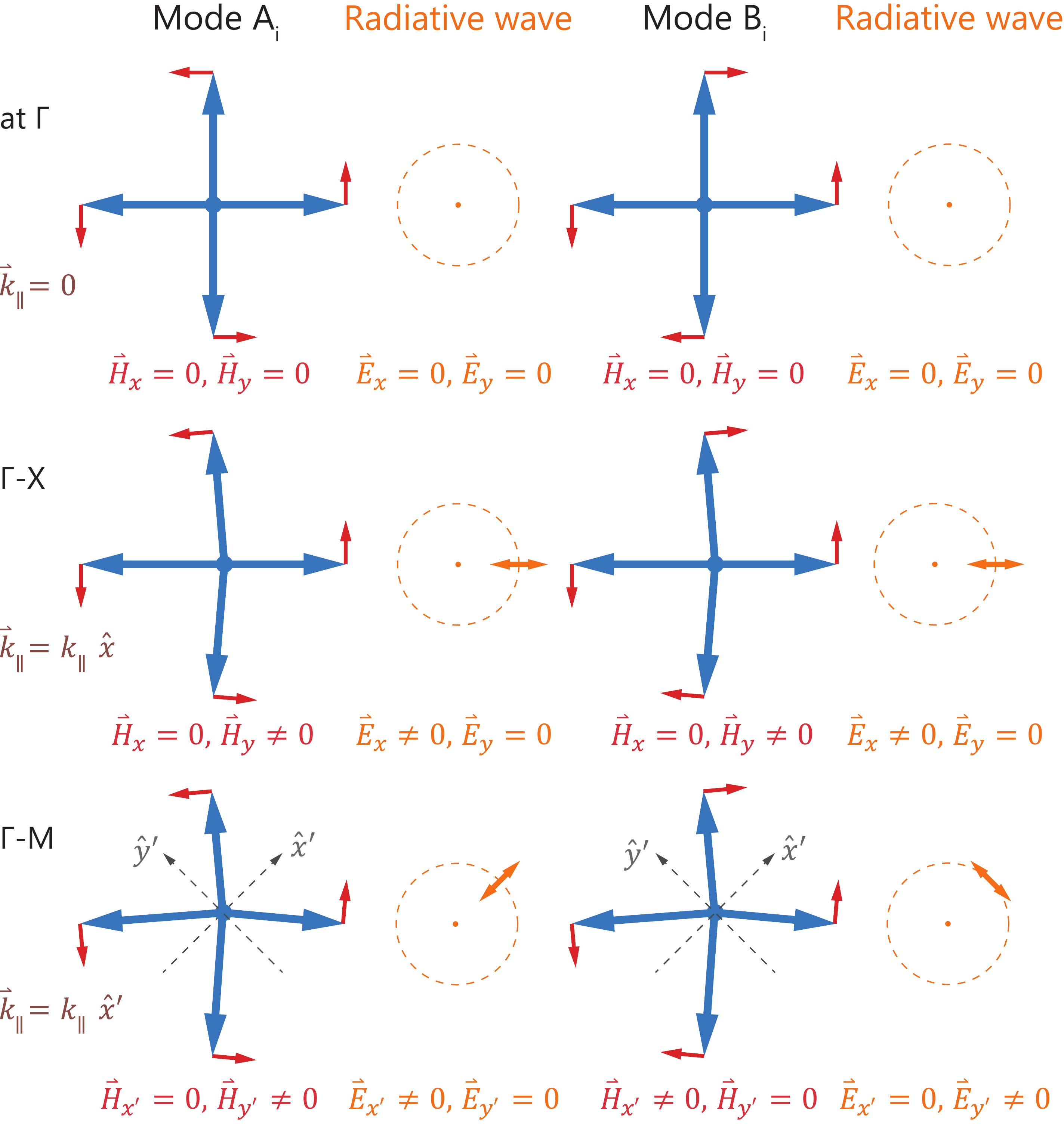}
\caption{Symmetry and polarization analysis of TM modes $A_i$ and $B_i$ for an infinitely periodic photonic crystal structure. Here, three $k_{\parallel}$ points are considered in the Brillouin zone: at $\Gamma$, on $\Gamma-X$, and on $\Gamma-M$. The blue arrows indicate the wavevectors of the fundamental Bloch waves in a photonic crystal. The red arrows perpendicular to them show the corresponding in-plane magnetic field component $H_{\parallel}$. The orange panels show the polarization of the radiative waves at each $k_{\parallel}$ point.}\label{fig:inf_pol}
\end{figure}

The polarization of photonic crystal modes can be predicted by symmetry analysis of Bloch wave interference in an infinitely periodic structure, as shown in Fig. \ref{fig:inf_pol} for the TM case.
Take mode $A_i$ as an example. At $\Gamma$ point, the counter-propagating Bloch waves destructively interfere with each other. Thus, the overall radiation is canceled. On $\Gamma-X$ axis with a non-zero $k_x$ wavevector, the amplitudes of the two Bloch waves propagating along $x$ and $-x$ directions are no longer the same. Their interference leads to a non-zero $H_y$ component. For the other two Bloch waves that propagate slightly off the $y$ axis, their interference cancels the $H_x$ components. Therefore, the overall remaining magnetic field of the radiative wave is along the $y$ axis, which generates $E_x$ components. If $k_{\parallel}$ is on the $\Gamma-M$ axis, we can rotate the coordinates and redefine the diagonal axes as $x'$ and $y'$. Along the diagonal direction where $k_{\parallel}$ is on the $x'$ axis, similar analysis tells the overall radiative wave possesses a non-zero $H_{y'}$, and the polarization of the radiating wave is along the $x'$ axis.

The polarization of finite-size photonic crystal modes $A$ and $B$ on the $x$, $y$ and the diagonal axes are in agreement with that of $A_i$ and $B_i$ predicted based on Fig. \ref{fig:inf_pol}. It should be noted that, here the symmetries of the envelope functions are not considered in the analysis.

\subsection{In-plane intensity distribution of all the six fundamental modes}
\label{supsec:intensity}
\begin{figure}[!htbp]
\centering\includegraphics[width=0.8\linewidth]{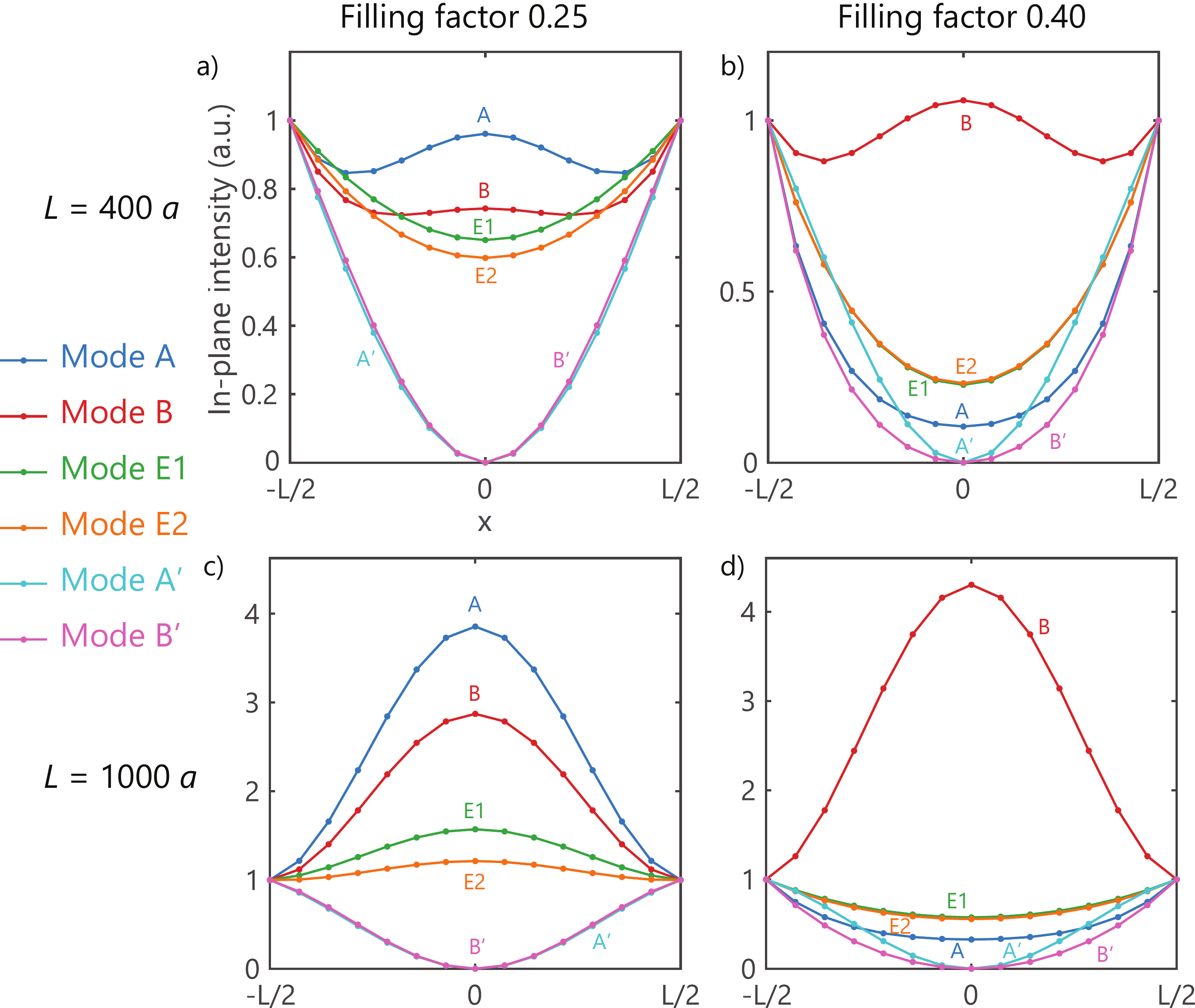}
\caption{Near-field in-plane intensity along the $x$ axis (through the device center point) with different pillar filling factors and device dimensions. The intensity profiles are calculated by the coupled wave theory for the six finite-size modes discussed in the main text.} \label{fig:intensity}
\end{figure}
Figure \ref{fig:intensity} shows the near-field intensity profiles of the six fundamental modes with different structural parameters.  When the filling factor is 0.25 and the device dimension is 400 $a$ ($a$ is the lattice-constant), the in-plane profiles of modes $A$, $B$ have anti-nodes in the device center, and modes $E1$ and $E2$ are concave in shapes. When the dimension is extended up to 1000 $a$, the profiles of modes $A$, $B$, $E1$ and $E2$ all rise into convex shapes with anti-nodes at the center, due to a higher in-plane confinement. At the filling factor of 0.40, modes $A$, $E1$, and $E2$ are in the vicinity of the triply-degenerate Dirac-like points \cite{zhen_spawning_2015,yin2017analytical}. They all exhibit leaky profiles with both 400 $a$ and 1000 $a$. In comparison, the in-plane intensities of modes $A'$ and $B'$ are concave in shapes and are pinned to zero at the device center in all cases.

\subsection{Envelope functions of all the six fundamental photonic crystal modes}
\label{supsec:envelope_all}
\begin{figure}[!htbp]
\centering\includegraphics[width=0.85\linewidth]{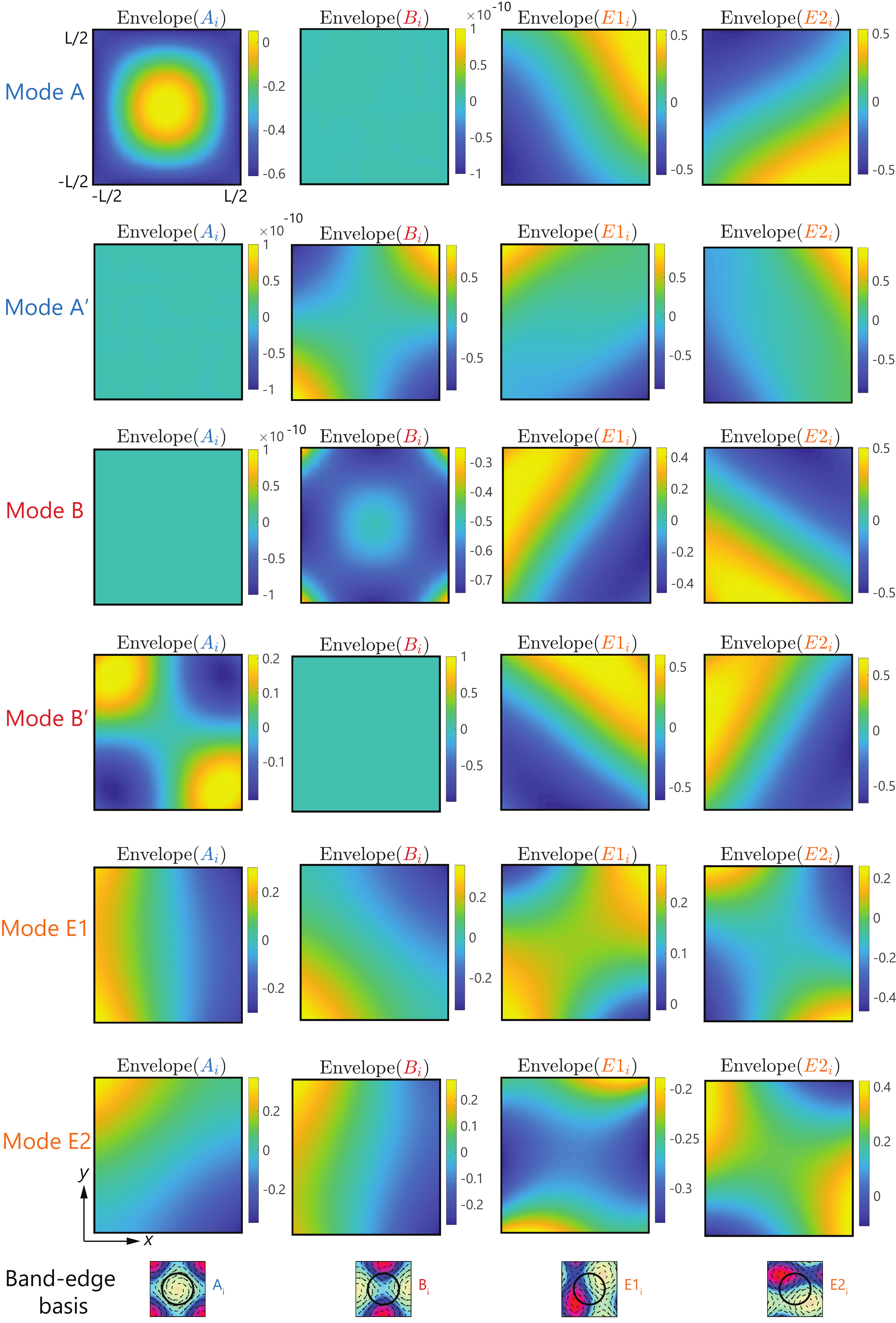}
\caption{Envelope functions of modes $A$, $A'$, $B$, $B'$, $E1$ and $E2$ on the basis of photonic crystal band-edge modes. The values are in arbitrary units. The device here has the area of $400 a \times 400 a$, where $a$ is the lattice-constant.} \label{fig:envs_all}
\end{figure}
Envelope functions of all the six photonic crystal modes $A$, $A'$, $B$, $B'$, $E1$ and $E2$ on band-edge basis are calculated according to Eq. 2 in the main text, and shown in Fig. \ref{fig:envs_all}. Part of this figure (modes $B$ and $B'$) is shown as Fig. 2 in the main text.

In the top row of Fig. \ref{fig:envs_all} (mode $A$), the envelope of the basis term $B_i$ has the maximum absolute amplitude that is no larger than $10^{-10}$, whereas the envelopes of other terms have a maximum absolute amplitude in the magnitude of $1$. Therefore, the envelope function of the term $B_i$ is considered as zero, which is also the case for the $A_i$ envelope of mode $A'$ and $B$, as well as the $B_i$ envelope of the mode $B'$.

\subsection{More details about the envelope analysis}
\label{supsec:envelope_analysis_addition}
\subsubsection{$C_4$ rotational symmetry of topological charge modes}
\begin{figure}[!htbp]
\centering\includegraphics[width=0.85\linewidth]{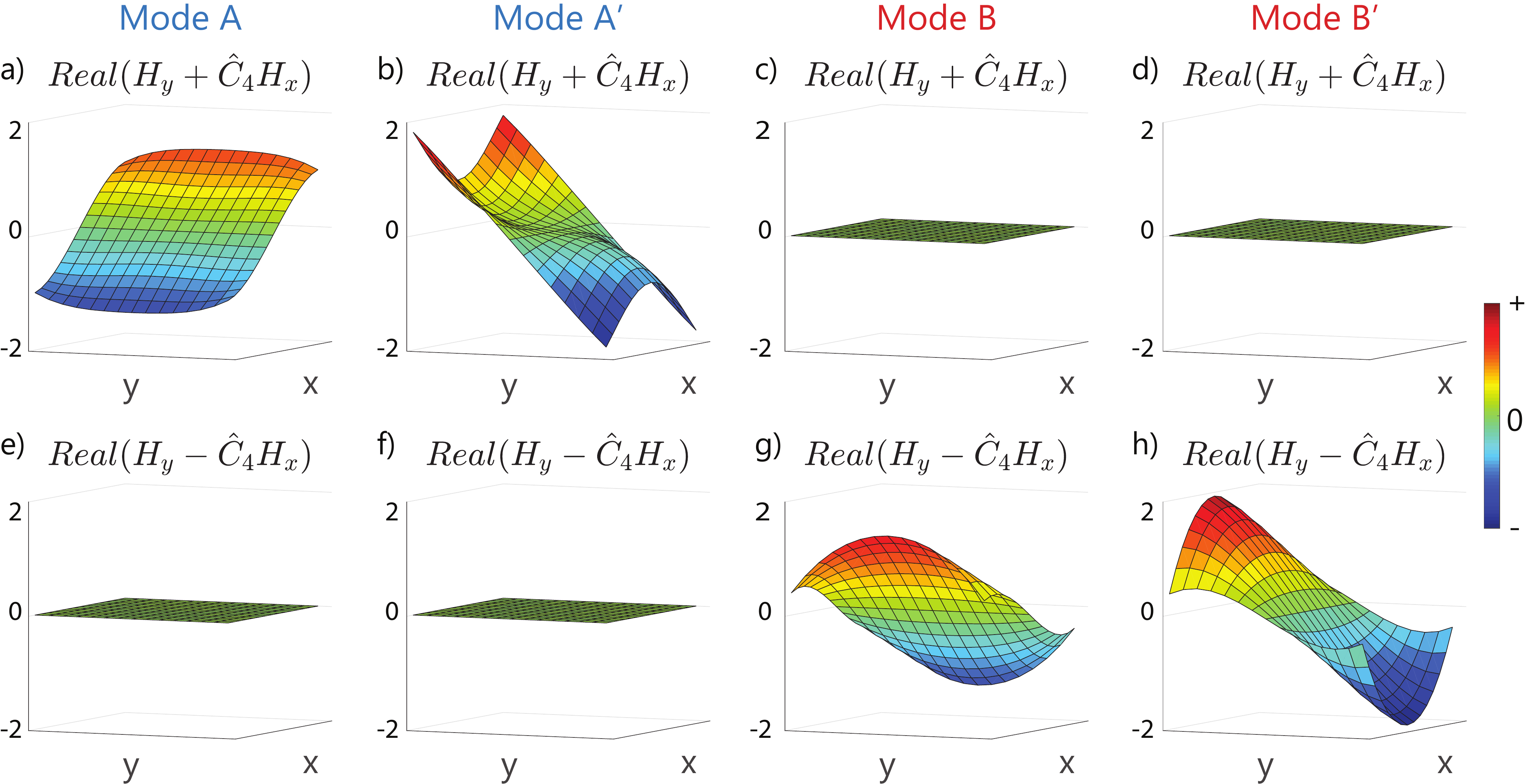}
\caption{$C_4$ rotational symmetry analysis of modes $A$ (a, e), $A'$ (b, f), $B$ (c, g) and $B'$ (d, h) at the filling factor of 0.25. $H_x$ and $H_y$ are the in-plane magnetic field components of the fundamental Bloch waves in the photonic crystal. (e, f) show the fields of modes $A, A'$ are $C_4$-even, and (c, d) show the fields of modws $B, B'$ are $C_4$-odd. The vertical axes are in arbitrary units.} \label{fig:ABOO}
\end{figure}

Figure \ref{fig:ABOO} shows the rotational symmetry properties of the in-plane magnetic field for modes $A$, $A'$, $B$ and $B'$, where $\hat{C}_{4}$ is a counter-clockwise 90$^{\circ}$ rotation operator. It proves that the fields of modes $A$ and $A'$ are even under a $C_4$ operation, whereas the fields of modes $B$ and $B'$ are odd under a $C_4$ operation. This explains values of their topological charges, and agrees with the results of the envelope analysis as Tab. I in the main text.


\subsubsection{Polarization analysis of mode $B'$}
\label{supsec:finite_pol}

\begin{figure}[!htbp]
\centering\includegraphics[width=0.85\linewidth]{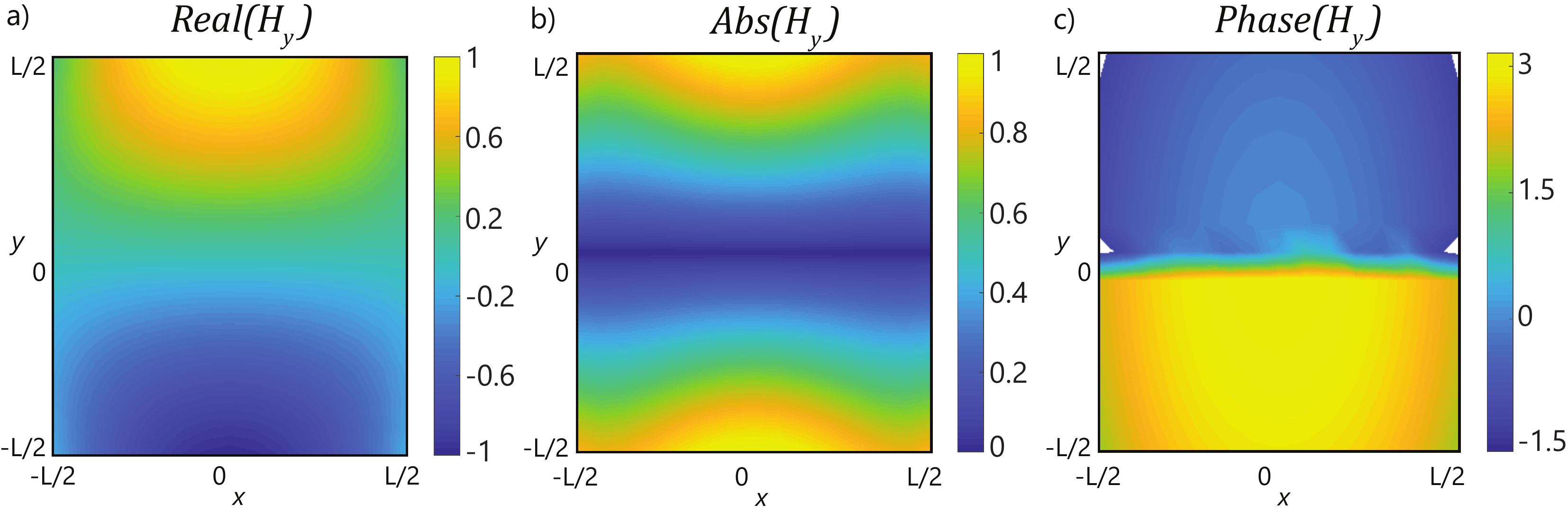}
\caption{Symmetry and polarization analysis based on the envelopes of the Bloch waves $R_x$, $S_x$ and their summation for the exceptional mode $B'$. The envelopes are calculated with our coupled wave theory model.}\label{fig:modeO}
\end{figure}

To understand the polarization property of mode $B'$, we show in Fig. \ref{fig:modeO} the near-field in-plane profile of the $H_y$ component. Along the $x$ axis ($y = 0$), the contributions from the field in the $y > 0$ region and the $y < 0$ region are in the same amplitude, but with the opposite phase. They destructively interfere, leading to zero $H_y$ components on the $x$ axis, which means the $E_x$ component of the radiating wave equals to zero. On the contrary, $E_x$ component is non-zero along the $y$ axis ($x = 0$), due to a constructive interference between the field in the $x > 0$ region and the $x < 0$ region. By analyzing the $H_x$ ($E_y$) component in the same manner, we can see that the polarization of mode $B'$ is azimuthal on the $x$ and $y$ axes. The polarization in the diagonal directions can be predicted in a similar manner, by analyzing the near-field envelopes of the $H_x + H_y$  and $H_x - H_y$ components. This detailed analysis further confirms the polarization of mode $B'$ as presented in Fig. 1(c) in the main text.

\subsection{Cavity losses of the modes}
\label{supsec:threshold}

\begin{figure}[!htbp]
\centering\includegraphics[width=0.75\linewidth]{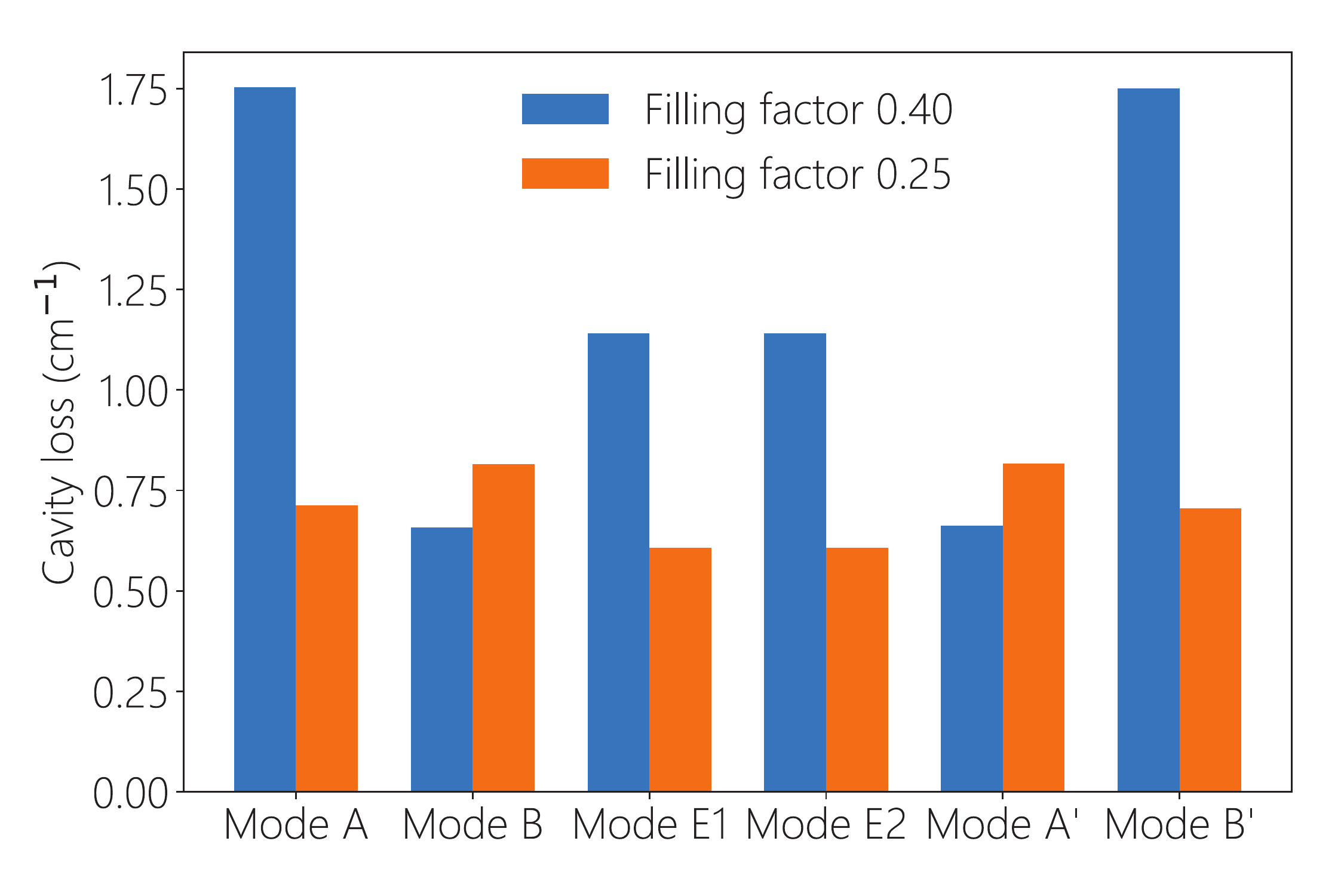}
\caption{The cavity loss of the six finite-size modes with filling factors of 0.40 and 0.25, calculated by coupled wave theory with the device area of $400 a \times 400 a$. }\label{fig:threshold}
\end{figure}

With the coupled wave theory, the mode cavity loss can be obtained by solving Eq. 1 in the main text. The results for the six fundamental finite-size modes on a $400 a \times 400 a$ structure are calculated with two pillar filling factors, as shown in Fig. \ref{fig:threshold}. At the pillar filling factor of 0.40, mode $B$ has the smallest cavity loss. At the pillar filling factor of 0.25, the cavity loss of mode $B'$ is slightly higher than that of modes $E1$ and $E2$, but mode $B'$ experiences a higher net gain compared to them. This is due to the non-uniform in-plane current injection \cite{liang2019room,wang2019large}, resulting from the large electrode window and the low doping of the substrate ($1.5 \times 10^{16}$ cm$^{-3}$). Therefore, we target mode $B'$ as the lasing mode for device II.

%


\end{document}